\documentclass[prb,twocolumn,showpacs,groupedaddress,floatfix]{revtex4}
\usepackage[dvips]{epsfig}
\usepackage{subfigure}
\usepackage{float}
\usepackage{amsmath}
\usepackage{amssymb}
\usepackage{amsfonts}
\usepackage{rotating}
\usepackage{euscript}
\usepackage{subfigure}
\usepackage{enumerate}
\usepackage{hhline}
\usepackage{threeparttable}
\usepackage{supertabular}
\usepackage{pslatex}
\usepackage{tabularx}
\usepackage{color}

\definecolor{blue}{rgb}{0,0.39,1}
\definecolor{red}{rgb}{1,0,0}
\definecolor{green}{rgb}{0,1,0}
\definecolor{cyan}{rgb}{0,1,1}
\definecolor{magenta}{rgb}{1,0,1}

\begin{document}

\title{\bf{Single quantum dot spectroscopy using a fiber taper waveguide near-field optic}}

\author{Kartik Srinivasan}
\email{phone: (626) 395-6269, fax: (626) 795-7258, e-mail:
kartik@caltech.edu} \affiliation{Center for the Physics of
Information and Department of Applied Physics, California
Institute of Technology, Pasadena, CA 91125}
\author{Andreas Stintz}
\author{Sanjay Krishna}
\affiliation{Center for High Technology Materials, University of
New Mexico, Albuquerque, NM 87106, USA.}
\author{Oskar Painter}
\affiliation{Center for the Physics of Information and Department
of Applied Physics, California Institute of Technology, Pasadena,
CA 91125}

\date{\today}

\begin{abstract}

  Photoluminescence spectroscopy of single InAs quantum dots at
  cryogenic temperatures ($\sim$ 14 K) is performed using a
  micron-scale optical fiber taper waveguide as a near-field optic.
  The measured collection efficiency of quantum dot spontaneous
  emission into the fundamental guided mode of the fiber taper is
  estimated at $0.1\%$, and spatially-resolved measurements with
  $\sim$600 nm resolution are obtained by varying the taper position with respect
  to the sample and using the fiber taper for both the pump and collection channels.

\end{abstract}
\maketitle

\noindent

Semiconductor quantum dots (QDs) have drawn significant
recent interest due to their potential to create novel optoelectronic
devices such as those that can generate single photons on demand
\cite{ref:Michler}. Methods to optically interrogate single QDs
include physical isolation via microfabricated mesas 
\cite{ref:Marzin} and diffraction-limited confocal optical
microscopy \cite{ref:Dekel}. Despite the technical challenges
involved in developing tools for use at cryogenic temperatures (a
prerequisite for most single QD studies), solid immersion lenses 
\cite{ref:Wu} have been implemented to improve the efficiency of
light collection, while near-field scanning optical microscopy
(NSOM) has been used to achieve sub-100 nm spatial resolution
\cite{ref:Toda}. In this paper, we examine the use of optical
fiber taper waveguides as a near-field optic for performing single
QD spectroscopy. These micron-scale silica waveguides have been
used in many studies of optical microcavities, beginning as an
efficient coupler to silica microspheres \cite{ref:Knight}. More
recently, we have shown that they can effectively probe the
spatial and spectral properties of small mode volume
($V_{\text{eff}}$), high refractive index semiconductor cavities
such as planar photonic crystals and microdisks
\cite{ref:Srinivasan7}. Other researchers have
proposed\cite{ref:Klimov1,ref:Kien1} and realized their use as a
collection tool for spontaneous emission from atomic vapors
\cite{ref:Nayak}. Here, we show that a fiber taper may be used to
channel emission from single self-assembled QDs embedded in a
semiconductor slab directly into a standard single-mode fiber with
high efficiency ($\sim 0.1\%$), and to provide sub-micron spatial
resolution of QDs, either through taper positioning or resonant
pumping of the optical modes of etched microdisk structures.

\begin{figure}[t]
\centerline{\includegraphics[width=8.3cm]{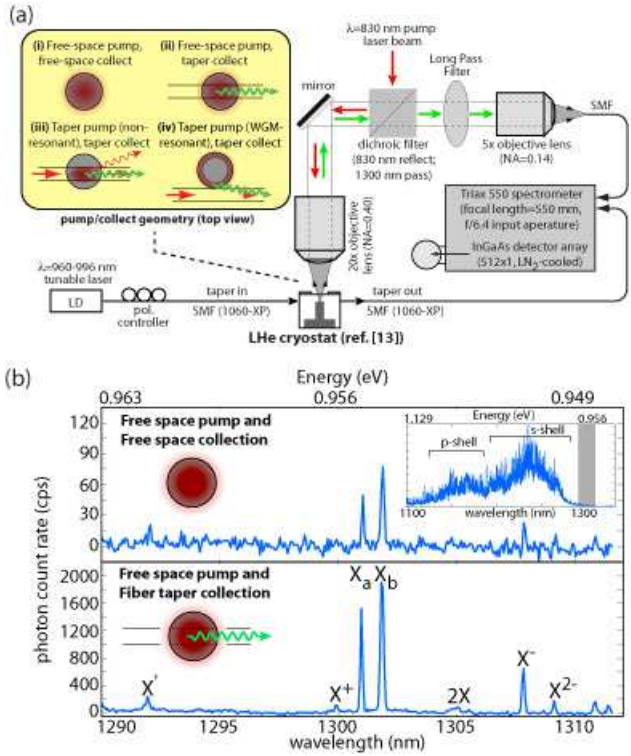}}
 \caption{(a) Schematic of the experimental apparatus, showing the pump and
   collection configurations studied here. (b) Emission from a single quantum dot (QD) using free-space collection (top) and
   fiber taper collection (bottom), under identical free-space pumping conditions ($\lambda_{p} = 830$ nm, $15.7$ nW
   incident power). The inset shows emission over a broad wavelength range from the ensemble of QDs within the
   microdisk. The shaded region from $\lambda=1290$-$1310$ nm is the spectral region where single QD emission has been
   observed.}\label{fig:PL_broad_scan_plus_fs_vs_fiber}
\end{figure}

The QDs we study consist of a single layer of InAs QDs embedded in
an In$_{0.15}$Ga$_{0.85}$As quantum well, a so-called
dot-in-a-well (DWELL) structure\cite{ref:Liu_G}.  The DWELL layer
is grown in the center of a GaAs waveguide (total waveguide
thickness of 256 nm), which sits atop a 1.5 $\mu$m thick
Al$_{0.7}$Ga$_{0.3}$As buffer layer.  The resulting peak of the
ground state emission of the ensemble of QDs is located at
$\lambda=1.35$ $\mu$m at room temperature. To limit the number of
optically pumped QDs, microdisk cavities of diameter $D=2$ $\mu$m
were fabricated using electron beam lithography and a series of
dry and wet etching steps \cite{ref:Srinivasan9}. Although the QDs
physically reside in a microcavity, they are
$\textit{non-resonant}$ with the cavity whispering gallery modes
(WGMs). In other words, our primary interest here is general
single QD spectroscopy through the fiber taper, without
enhancement through interaction with the high quality factor ($Q$)
microdisk WGMs. The samples were mounted in a continuous-flow
liquid He cryostat that has been modified to allow sample probing
with optical fiber tapers while being held at cryogenic
temperatures (T$\sim 14$ K), as described in detail in ref.
[\onlinecite{ref:Srinivasan14}]. The cryostat is part of a
microphotoluminescence setup that provides any combination of
free-space and fiber taper pumping and collection; see Fig.
\ref{fig:PL_broad_scan_plus_fs_vs_fiber}(a) for details.

The inset of Fig. \ref{fig:PL_broad_scan_plus_fs_vs_fiber}(b)
shows the emission spectrum from an ensemble of QDs in one of the
microdisks. Here, the device is optically pumped through an
objective lens at normal incidence (free-space pumping), with a
spot size of $3$ $\mu$m and wavelength $\lambda_{P}=830$ nm.
Clearly present are the ground and excited states ($s$ and $p$
shells) of the ensemble of QDs which, based on the estimated QD
density of 300-500 $\mu$m$^{-2}$, consists of $\sim 1000$ QDs. To
study isolated emission lines from single QDs, we focus on the
long-wavelength tail end of the QD distribution
($\lambda=1290$-$1310$ nm). In this range, isolated emission lines
from a single QD are seen for a fraction ($10\%$) of the
interrogated devices.  A typical spectrum as collected through the pump lens (free-space collection) from one such
device is shown in the top panel of Fig.
\ref{fig:PL_broad_scan_plus_fs_vs_fiber}(b). Under identical
pumping conditions, the signal collected through a fiber taper
waveguide positioned on top of, and in contact with, the microdisk
is shown in the bottom panel of Fig.
\ref{fig:PL_broad_scan_plus_fs_vs_fiber}(b). The taper is a single
mode optical fiber that has been heated and stretched down to a
minimum radius of $a = 650$ nm, and is installed in the customized
liquid He cryostat as detailed in ref.
[\onlinecite{ref:Srinivasan14}].  The most stark difference
between the free-space and fiber taper collected spectra is the
$25 \times$ increase in fiber taper collected power.  Similar
improvement in collection efficiency was measured over all the QDs
studied in this work.

\begin{figure}
\centerline{\includegraphics[width=8.3cm]{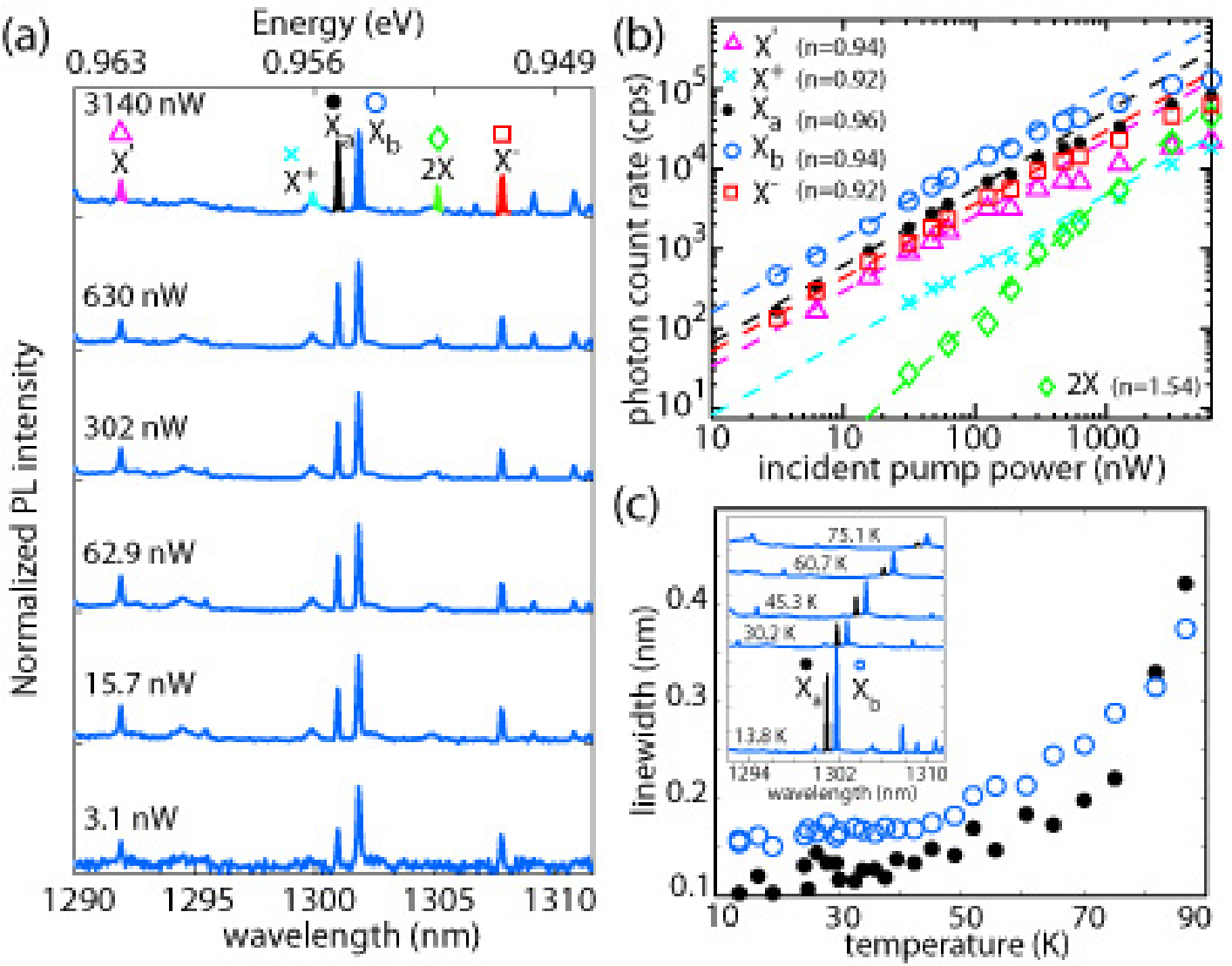}}
 \caption{(a) Normalized emission spectra from a single QD for different free-space
   incident pump powers and using fiber taper collection.  (b) Log-log plot of the collected emission for the $X_{a}$
   ($\textcolor{black}{\bullet}$), $X_{b}$ ($\textcolor{blue}{\circ}$), $X^{+}$ ($\textcolor{cyan}{\times}$), $X^{-}$
   ($\textcolor{red}{\Box}$), $2X$ ($\textcolor{green}{\Diamond}$), and $X^{\prime}$ ($\textcolor{magenta}{\triangle}$)
   QD states of part (a).  The dashed lines are least squares fits to the emission data below saturation, assuming $I
   \sim P^{n}$, where $I$ is the collected photon count rate emitted into a given line and $P$ is the pump power. (c)
   Linewidth of the QD exciton states $X_{a}$ and $X_{b}$ as a function of temperature.  The inset shows representative
   spectra at different temperatures.}
\label{fig:pump_power_dependence}
\end{figure}

Before further studying the fiber taper as a collection optic, we
attempt to identify the different QD lines of Fig.
\ref{fig:PL_broad_scan_plus_fs_vs_fiber}(b). Of particular benefit
in this assignment is the recent work of Cade, $\textit{et al.}$
\cite{ref:Cade1}, who study a DWELL material very similar to that
investigated here. In Fig. \ref{fig:pump_power_dependence}(a), we
show taper-collected emission spectra as a function of pump power
(free-space, $\lambda_{P}=830$ nm) for a fixed taper position.
Emission is first seen for incident powers of a few nW (estimated
absorbed powers of tens of pW), with excitonic lines centered at
$1291.95$ nm, $1300.97$ nm, $1301.81$ nm, and $1307.75$ nm. As we
discuss later, spatially-resolved measurements clearly indicate
that the shortest wavelength emission line is unrelated to the
latter three, which we identify as the polarization-split exciton
lines ($X_a$ and $X_b$) \cite{ref:Cade1,ref:Bayer2} and the
negatively charged exciton line ($X^{-}$). As the pump power is
increased, additional emission lines appear, including the
positively charged exciton ($X^{+}$) at $1299.87$ nm and the
bi-exciton ($2X$) line at $1305.11$ nm. The $X^{-}$-$X$,
$X^{+}$-$X$, and $2X$-$X$ splitting values of $4.6$, $-1.1$, and
$2.8$ meV match reasonably well with the $5.6$, $-1.1$, and $3.1$
meV values measured in Ref.  [\onlinecite{ref:Cade1}], although
the fine structure splitting in the $X$ line is significantly
larger ($600$ vs. $300$ $\mu$eV) for this QD. In Fig.
\ref{fig:pump_power_dependence}(b), we plot the emission level in
each QD state against pump power. Below saturation, the emission
lines all scale nearly linearly with pump power, except for the
$2X$ line which scales superlinearly, although more slowly than
expected ($n=1.54$ as opposed to $2$).  Previous studies of
$1.2$-$1.3$ $\mu$m QDs have also measured a sub-quadratic
pump-power-dependence for the $2X$ line
\cite{ref:Cade1,ref:Alloing},although usually in conjunction with
a sub-linear dependence of the $X$ line. Finally, the temperature
(T) dependence of the $X$ lines is shown in Fig.
\ref{fig:pump_power_dependence}(c), where significant broadening
is seen for T$>50$ K.  Below this temperature we measure
linewidths of $0.1$-$0.15$ nm, roughly corresponding to the
spectral resolution of our system ($0.1$ nm $= 75$ $\mu$eV).

A rough estimate of the absolute collection efficiency of the
fiber taper is derived by considering the saturated photon count
rates for the $X$ lines in Fig.
\ref{fig:pump_power_dependence}(b). The measured saturated photon
count rate into the $X_{b}$ line is $1.5 \times 10^{5}$ counts per
second (cps), which after considering the spectrometer
grating efficiency ($60\%$), the detector array quantum efficiency
($85\%$), and including the light in the backwards fiber channel,
corresponds to a count rate of $5.9 \times 10^{5}$ cps.  Taking
into account the transmission effiency of the fiber taper
($\sqrt{0.64}$ for a QD centrally located along the tapered region
of the fiber), the taper-collected photon count rate rises to $7.4
\times 10^{5}$ cps. Neglecting possible suppression or enhancement
of radiation due to the presence of the microdisk (a good
approximation for QDs located above the disk's central pedestal),
the saturated photon emission rate for InAs QDs is $5$-$10 \times
10^{8}$ cps (photon lifetime $\tau=1$-$2$ ns\cite{ref:Michler}).
This yields an approximate fiber taper collection efficiency of
$\eta_{t}=0.1\%$. It is important to note that this efficiency is
for non-resonant collection, and does not correspond to that
attainable for QDs resonant with a high-$Q$ microdisk WGM, which
one would expect to be much higher due to the Purcell-enhanced
emission into a localized cavity mode\cite{ref:Michler} and the
efficient taper-WGM coupling\cite{ref:Srinivasan9}.

Obtaining a theoretical value for the non-resonant fiber taper
collection efficiency is hindered by the complex geometry in which
the QD is embedded; however, a coarse estimate can be made by
comparison to refs. [\onlinecite{ref:Klimov1,ref:Kien1}]. In these
works, a spontaneous emission collection efficiency of $20$-$50\%$
is estimated for a dipole emitter on the surface of a silica fiber
taper of radius $a \sim 1.4/k_{0}$ ($k_{0}$ is the free-space
wavenumber). For our larger fiber taper ($a \sim 3/k_{0}$), and
for a QD $125$ nm away from its surface (corresponding to the
middle of the GaAs slab), the theoretical collection efficiency is
on the order of $1\%$. Beyond providing an upper bound on the
collection efficiency (due to the high-index GaAs slab and AlGaAs
substrate of the measured devices), the model indicates that we
are far from the optimum fiber taper diameter. Comparison with
ref. [\onlinecite{ref:Kien1}] indicates that an order of magnitude
increase in the collection efficiency may be obtained by
decreasing the fiber taper radius to $a=300$ nm, an experimentally
realizable value.

\begin{figure}[t]
  \centerline{\includegraphics[width=8.3cm]{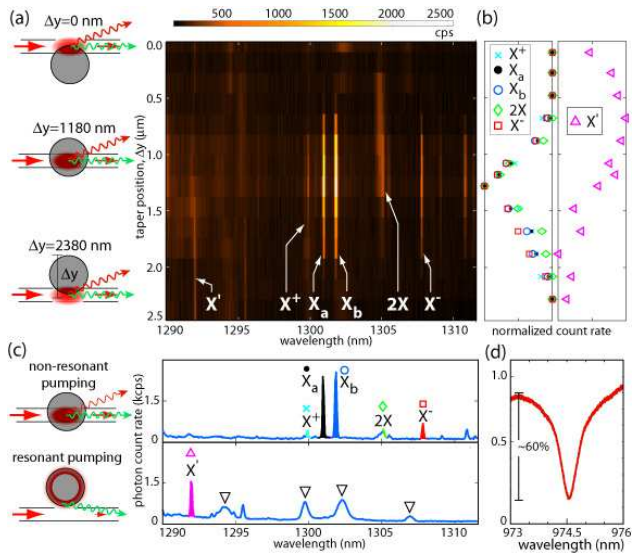}}
 \caption{Spatially-resolved measurements using fiber taper pumping and
   collection. (a) Collected emission spectrum as a function of taper position along the $\hat{y}$-axis of the sample,
   for non-resonant pumping through the fiber taper. (b) Spatial dependence of the integrated emission into each QD
   line.  (c) Emission spectrum for non-resonant pumping with the taper at ${\Delta}y$=1180 nm and an input power of
   $125$ nW (top), and for resonant pumping of a microdisk WGM in the $\lambda_{P}=980$ nm pump band at a power of $690$
   pW (bottom).  The triangles ($\textcolor{black}{\bigtriangledown}$) indicate the position of WGM resonances.  (d)
   Wavelength scan of the WGM used to pump the sample in (c).}
 \label{fig:spatial_dependence}
\end{figure}

An additional benefit of using the fiber taper as a near-field
collection optic is the potential for spatially-resolved
measurements. Although the spatial resolution provided by a glass
fiber taper \cite{ref:Srinivasan7} is lower than the sub-100 nm
level achievable through NSOM \cite{ref:Toda}, valuable
information on the spatial location of QDs can be inferred from
both the spatially-dependent collection and excitation through the
fiber taper. For the following measurements, we pump the microdisk
through the fiber taper (instead of from free-space), with
$\lambda_{p}=978.3$ nm, where only the DWELL is significantly
absorbing. Figure \ref{fig:spatial_dependence}(a) shows a plot of
the fiber-collected emission spectrum as a function of taper
position along the $\hat{y}$-axis of the sample (the taper
position is adjusted through a piezo stage on which it is
mounted). Figure \ref{fig:spatial_dependence}(b) plots the spatial
dependence of the total photon count rate within each of the QD
states identified in Fig. \ref{fig:pump_power_dependence}. Of note
is the similar spatial dependence of the collected emission from
each of the lines $\{X_{a},X_{b},X^{+},X^{-},2X\}$, confirming
that they originate from the same single QD.  The full-width at
half-maximum of the collected emission is roughly 600 nm, giving
an estimate of the taper's spatial resolution transverse to its
longitudinal axis.  Two-dimensional mapping of a QD's position may
also be obtained by rotating the sample, and repeating the
measurement along the orthogonal axis as was done in ref.
[\onlinecite{ref:Srinivasan7}] for the mapping of the modes of
photonic crystal cavities.  Finally, in contrast to the other
emission lines, emission from the short wavelength $X^{\prime}$
line at $1292$ nm has a quite different dependence on taper
position, with collected emission being strongest for the taper at
the disk periphery.

Spatial selection of QDs may also be realized by resonantly
pumping a microdisk WGM.  This excites QDs located in a 250 nm
thick annulus at the microdisk perimeter, where the pump beam
resides, and efficient taper-WGM coupling allows for an accurate
estimate of the \emph{absorbed} pump power. The QDs located at the
disk periphery are of course those that are of interest for cavity
QED studies involving high-$Q$, ultrasmall $V_{\text{eff}}$ WGMs.
Figure \ref{fig:spatial_dependence}(d) shows a transmission scan
of a pump-band WGM with a coupling depth of $60\%$ and
$Q{\sim}$1000 (limited by DWELL absorption). By pumping on
resonance at $\lambda_{p}=974.5$ nm, we reduce the power needed to
achieve a given signal by $2$-$3$ orders of magnitude relative to
non-resonant pumping. The bottom scan of Fig.
\ref{fig:spatial_dependence}(c) shows the emission spectrum when
we pump on resonance with 690 pW of power at the taper input
(corresponding to $330$ pW of dropped/absorbed power). Emission
from the centrally located QD (top scan of Fig.
\ref{fig:spatial_dependence}(c)) is no longer present, and has
been replaced by a pronounced emission peak at $\lambda=1291.95$
nm, corresponding to the $\text{X}^{\prime}$ line, confirming that
this emission is likely due to a QD located in the disk periphery.
Another difference in comparison to the non-resonant pumping
spectrum is the presence of several broad emission peaks.  These
peaks are due to emission into relatively low-$Q$,
higher-radial-order WGMs of the microdisk, as confirmed by
fiber-taper-based transmission spectroscopy of the cavity with a
tunable laser \cite{ref:Srinivasan9}.  The source of such
background emission into detuned cavity modes is not well
understood, but has been observed to occur for even large
detunings of $10$-$20$ nm \cite{ref:Hennessy3}. In this case, it
is likely that the preferential excitation of QDs that reside in
the microdisk perimeter, even those that have exciton lines which
are significantly detuned spectrally, results in enhanced emission
into the microdisk WGMs.

In summary, we have shown that a micron-scale optical fiber taper
waveguide, previously demonstrated to be an effective tool for
characterization of semiconductor microcavities, can also be used
to study single semiconductor quantum dots. As a near-field
collection optic, the fiber taper is shown to channel quantum dot
light emission directly into a single mode fiber with a high
efficiency of $0.1\%$, and to provide a sub-micron spatial
resolution of QDs.  The ability to effectively investigate both
microcavities and quantum dots suggests that these fiber tapers
can serve as a very versatile tool for studying microphotonic
structures, and in particular, for investigations of chip-based
cavity QED.

\bibliography{./PBG_4_30_2007}

\end{document}